\documentclass[aps,prb,twocolumn,amsmath,amssymb,floatfix,showpacs]{revtex4}
\usepackage{graphicx}
\bibstyle{apsrev.bib}

\begin{document}

\title{Recursive calculation of the microcanonical density of states}

\author{Lo\"{\i}c Turban}

\email{Loic.Turban@univ-lorraine.fr}


\affiliation{Groupe de Physique Statistique, D\'epartement Physique de la Mati\`ere et des Mat\'eriaux,
Institut Jean Lamour\footnote[2]{Laboratoire associ\'e au CNRS UMR 7198.}, CNRS---Universit\'e de Lorraine,\\
BP 70239, F-54506 Vand\oe uvre l\`es Nancy Cedex, France}


\begin{abstract}
For a   classical system of noninteracting particles we establish recursive integral equations for the density of states on the microcanonical ensemble. The recursion can be either on the number of particles or on the dimension of the system. The solution  of the integral equations is particularly simple when the single-particle density of states in one dimension follows a power law. Otherwise it can be obtained using a Laplace transform method. Since the Laplace transform of the microcanonical density of states is the canonical partition function, it factorizes for a system of noninteracting particles and the solution of the problem is straightforward. The results are illustrated on several classical  examples. 
\end{abstract}


\maketitle

\section{Introduction}
In statistical mechanics, an isolated macroscopic system at equilibrium is characterized by its microcanonical entropy,  given by the Boltzmann formula
\begin{equation}
S=k_{\mathrm B}\ln \Omega\,,
\label{S}
\end{equation}
where $k_{\mathrm B}$ is the Boltzmann constant and $\Omega$ the number of accessible microstates in phase space, i.e., the number of equiprobable microstates compatible with the fixed values of the macroscopic external parameters such as the volume $V$, the number of particles $N$, the energy $E$, etc. 
For a classical system in $D$ dimensions, the phase space has  $2s$ dimensions corresponding to the components of the generalized momenta $\mathbf  p$ and coordinates $\mathbf  q$ such that $s=DN$. Each microstate corresponds to a $2s$-dimensional cell in phase space with a volume $h^s$, where $h$ is the Planck constant. The accessible volume is such that 
\begin{equation}
E<H({\mathbf  p},{\mathbf  q})<E+\delta E\,,
\label{E}
\end{equation}
where the arbitrary uncertainty $\delta E\ll E$ affecting the value of the total energy is introduced in order  to obtain a $2s$ dimensional accessible volume, thus leading to a nonvanishing value of $\Omega$.

The number of accessible microstates is proportional to $\delta E$ and can be written as
\begin{equation}
\Omega=n_N(E)\delta E\,,
\label{def-dens}
\end{equation}
where the coefficient of proportionality $n_N(E)$ is the microcanonical density of states which plays a central role in the statistical physics of isolated systems. 

The calculation of the density of states  amounts to determine the volume enclosed by the  hypersurface of constant energy $E=H({\mathbf  p},{\mathbf  q})$ in the $2s$ dimensional phase space, a quite difficult task in general.  In  textbooks the properties of the microcanonical ensemble are often  illustrated by the study of some  classical systems involving noninteracting particles for which the Hamiltonian is separable and the constant energy surface simple enough.  In the case of the ideal gas one needs the area  of a hypersphere and for a collection of harmonic oscillators the area  of a hyperellipsoid which can be transformed to a hypersphere through rescaling. The  area of the hypersphere is usually obtained by comparing the values, calculated either in Cartesian or in spherical coordinates, of the integral over  the $2s$-dimensional space of $\exp(-r^2)$, where ${\mathbf r}$ is a radius vector (see,  e.g., Refs~\onlinecite{kubo67a,huang87,diu89,garrod95,mazenko00}). Other independent particle 
problems have also been treated\cite{fernandez-pineda79,roman95} by making use of Dirichlet's integral formula.\cite{whittaker27}

The purpose of the present paper is to  propose another  approach to noninteracting particle systems, based on a recursion relation for the microcanonical density of states.  Assuming that the single-particle density of states is known, the particles are added one by one. Summing their contribution to the total density of  states, a recursive integral equation is obtained which can be solved using a Laplace transform method. When the single-particle Hamiltonian is itself separable, involving a sum of  contributions from each of the $D$ dimensions, the same recipe can be used to evaluate the single-particle density of states in $D$ dimensions.

The recursive approach to the calculation of the $N$-particle microcanonical density of states is presented in Sec. II. Some illustrative examples are discussed in Sec. III. Summary and conclusion are given in Sec. IV.

\section{Microcanonical  density of states}

We consider an isolated classical $D$-dimensional system containing  $N$ noninteracting distinguishable particles. Let $N_p(E)$ denote the $p$-particle integrated density of states, i.e., the number of microstates with energy smaller than $E\geq0$ in a system with $p$ particles. The $p$-particle density of states, $n_p(E)$, is such that $dN_p(E)=n_p(E)dE$.  

The number  of microstates with energy smaller than $E$ in a system with $p+1$ particles is obtained by first combining the microstates with energy smaller than $E-\epsilon$ in a system with $p$ particles with the microstates  on the energy interval $[\epsilon,\epsilon+d\epsilon]$ associated with 
the added particle and then integrating over $\epsilon$. Doing so,  one obtains
\begin{equation}
N_{p+1}(E)=\int_0^E N_p(E-\epsilon)n_1(\epsilon)d \epsilon\,.
\label{rec-int-dens}
\end{equation}
Taking the derivative of both sides and noticing that $N_p(0)=0$,  leads to a recursion relation for the $p$-particle density of states
\begin{equation}
n_{p+1}(E)=\frac{dN_{p+1}(E)}{d E}=\int_0^E n_p(E-\epsilon)n_1(\epsilon)d \epsilon\,,
\label{rec-dens}
\end{equation}
The expression on the right is  a convolution product  thus, taking the Laplace transform of both sides,  one obtains
\begin{eqnarray}
z_{p+1}(s)&=&\int_0^\infty \! e^{-sE}n_{p+1}(E)\,dE\nonumber\\
&=&\int_0^\infty \! e^{-sE}dE \int_0^E \!  n_p(E-\epsilon)n_1(\epsilon)d \epsilon\nonumber\\
&=&\int_0^\infty \!   n_1(\epsilon) e^{-s\epsilon}d \epsilon\int_\epsilon^\infty \!    e^{-s(E-\epsilon)} n_p(E-\epsilon) dE\nonumber\\
&=&z_p(s)\,z_1(s)\,.
\label{conv}
\end{eqnarray}
Iterating this result,  the Laplace transform of the N-particle density of states is given by
\begin{equation}
z_N(s)=[z_1(s)]^N\,,
\label{conv-N}
\end{equation}
when the particles are identical. Otherwise the power is replaced by a product over the different contributions. The $N$-particle density of states is obtained by taking the inverse Laplace transform
\begin{equation}
n_N(E)=\frac{1}{2\pi i}\int_{\gamma-i\infty}^{\gamma+i\infty}\! e^{sE}z_N(s)\, ds\,,
\label{inv}
\end{equation}
where $\gamma$ is a vertical contour in the complex plane chosen so  that all singularities of $z_N(s)$ are on the left of it.

One may notice that with $1/s=k_{\mathrm B}T$, $ z_N(s)$ is the $N$-particle canonical partition function and Eq.~(\ref{conv-N}) simply states that   for noninteracting particles the $N$-particle canonical partition function factorizes and is given by the $N$th power of the single-particle canonical partition function $z_1(s)$.

Let us  assume that the single-particle density of states varies as  a power of the energy 
\begin{equation}
n_1(E)=C_1E^{\kappa_1-1}\,,\qquad E\geq0\,.
\label{dens-1}
\end{equation}
The Laplace transform is then given by
\begin{eqnarray}
z_1(s)&=&C_1\int_0^\infty \! e^{-sE}E^{\kappa_1-1}\, dE\nonumber\\
&=&\frac{C_1}{s^{\kappa_1}}\int_0^\infty \! e^{-t}t^{\kappa_1-1}\, dt=C_1\frac{\Gamma(\kappa_1)}{s^{\kappa_1}}\,,
\label{z1s}
\end{eqnarray}
where $\Gamma(x)$ is the Euler Gamma function.
For  the $N$-particle system we have
\begin{equation}
z_N(s)=\frac{[C_1\Gamma(\kappa_1)]^N}{s^{N\kappa_1}}\,.
\label{zNs}
\end{equation}
The inverse Laplace transform can be deduced from Eq.~(\ref{z1s}) and leads to the $N$-particle density of states
\begin{equation}
n_N(E)=[C_1\Gamma(\kappa_1)]^N\frac{E^{N\kappa_1-1}}{\Gamma(N\kappa_1)}\,.
\label{dens-N}
\end{equation}

In this simple case one may also proceed as follows: Using Eq.~(\ref{dens-1}) and the power-law expression of the $p$-particle density of states 
\begin{equation}
n_p(E)=C_pE^{\kappa_p-1}\,.
\label{dens-p}
\end{equation}
in  the integral equation~(\ref{rec-dens}) leads to 
\begin{eqnarray}
n_{p+1}(E)&=&C_1C_p\int_0^E \epsilon^{\kappa_1-1}(E-\epsilon)^{\kappa_p-1} d\epsilon\nonumber\\
&=&C_1C_pE^{\kappa_1+\kappa_p-1}\int_0^1t^{\kappa_1-1}(1-t)^{\kappa_p-1} dt\nonumber\\
&=&C_1C_p\frac{\Gamma(\kappa_1)\Gamma(\kappa_p)}{\Gamma(\kappa_1+\kappa_p)}
E^{\kappa_1+\kappa_p-1}\,.
\label{int-rec}
\end{eqnarray}
The integral in the second line is the Euler Beta function\cite{abramowitz72} which is written in terms of Gamma functions  in  the last  line.
Comparing with the form of $n_{p+1}(E)$ resulting from Eq.~(\ref{dens-p}) we deduce the following recursion relations for the amplitude and the exponent:
\begin{equation}
C_{p+1}=C_1C_p\frac{\Gamma(\kappa_1)\Gamma(\kappa_p)}{\Gamma(\kappa_1+\kappa_p)}\,,
\qquad \kappa_{p+1}=\kappa_1+\kappa_p\,.
\label{rec-c-om}
\end{equation}
Iterating these relations, one easily obtains
\begin{equation}
C_p=\frac{[C_1\Gamma(\kappa_1)]^p}{\Gamma(p\kappa_1)}\,,\qquad \kappa_p=p\kappa_1\,,
\label{rec-sol}
\end{equation}
and the form of the $N$-particle  density of states given in Eq.~(\ref{dens-N}) is recovered.

For a mixture with $N_a$ particles of type $a$ and $N_b$ particles of type $b$, we have 
a straightforward generalization  of Eq.~(\ref{dens-N}), namely
\begin{eqnarray}
&&\!\!\!n_{N_a,N_b}(E)=\int_0^E n_{N_a}(\epsilon)n_{N_b}(E-\epsilon)d\epsilon\ \ \ \ \ \ \ \ \ \ \ \ \ \ \ \ \ \ \ \ \ \ \nonumber\\
&&\ \ \ \ \ \ =[C_{a}\Gamma(\kappa_{a})]^{N_a}[C_{b}\Gamma(\kappa_{b})]^{N_b}
\frac{E^{N_a\kappa_{a}+N_b\kappa_{b}-1}}
{\Gamma(N_a\kappa_{a}+N_b\kappa_{b})}\,.
\label{dens-N-ab}
\end{eqnarray}

In order to calculate the microcanonical density of states, one needs the form of the single-particle density of states.

When the single-particle Hamiltonian is the sum of equivalent (kinetic and potential) contributions coming from each of the $D$ dimensions of the system, the number of dimensions plays the same role as the number of particles did previously and a recursion relation similar to Eq.~(\ref{rec-dens}) can be written  for the single-particle density of states in $q+1$ dimensions
\begin{equation}
n_1^{(q+1)}(E)=\int_0^E n_1^{(q)}(E-\epsilon)n_1^{(1)}(\epsilon)\, d \epsilon
\label{rec-dens-1}
\end{equation}
where the upper index now refers to the dimension of the system while the lower one refers  to the number of particles.

As before the convolution product leads to a form similar to~(\ref{conv-N}) 
\begin{equation}
z_1^{(D)}(s)=[z_1^{(1)}(s)]^D\,,
\label{conv-D}
\end{equation}
for the Laplace transform of the single-particle density of states in $D$ dimensions.  

When  the single-particle density of states in one dimension behaves as 
\begin{equation}
n_1^{(1)}(E)=C_1^{(1)}E^{\kappa_1^{(1)}-1}\,,\qquad E\geq0\,,
\label{dens-1-1}
\end{equation}
the single-particle density of states in $D$ dimensions  reads
\begin{equation}
n_1^{(D)}(E)=\left[C_1^{(1)}\Gamma(\kappa_1^{(1)})\right]^D
\frac{E^{D\kappa_1^{(1)}-1}}{\Gamma\left(D\kappa_1^{(1)}\right)}\,.
\label{dens-1-D}
\end{equation}
in analogy with  Eq.~(\ref{dens-N}).

When the system is anisotropic with $D_i$ dimensions  ($i=a,b$) for which the one-dimensional single-particle density of states is $n_1^{(i)}(E)$ with Laplace transform $z_1^{(i)}(s)$, Eq.~(\ref{conv-D}) has to be replaced by
\begin{equation}
z_1^{(D_a,D_b)}(s)=[z_1^{(a)}(s)]^{D_a}[z_1^{(b)}(s)]^{D_b}\,.
\label{conv-D-ab}
\end{equation}

\section{Some examples}

\subsection{Boltzmann  Ideal gas}

\begin{figure}[tbh]
\includegraphics[width=\columnwidth]{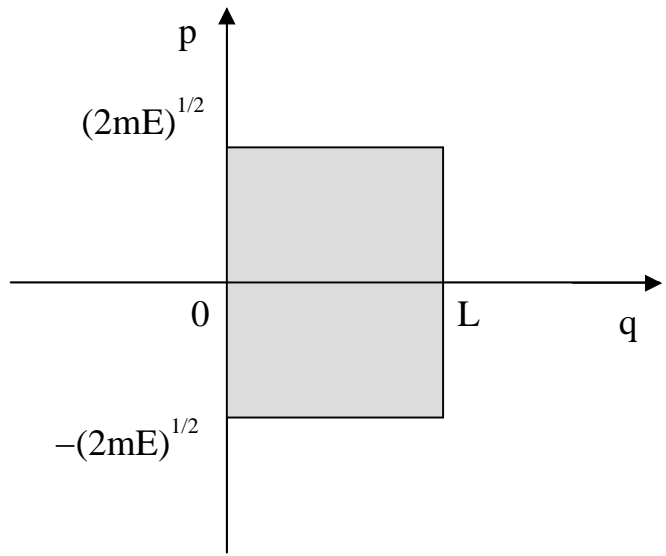}
\caption{Rectangular domain of accessible microstates with energy lower than $E$ in the phase space of a free particle with mass $m$,  confined on a segment with length $L$.}
\label{fig1}
\end{figure}

We consider an ideal gas of $N$ point particles of mass $m$, confined inside a cubic box  of volume $V_D=L^D$. 
The single-particle Hamiltonian is
\begin{equation}
H(p)=\frac{p^2}{2m}\,.
\label{H-ig}
\end{equation}
In the two-dimensional phase space of the single-particle  problem there are two segments of constant energy, $E$, such that $p(E)=\pm\sqrt{2mE}$ and $0<q<L$. Hence, as shown in Fig.~\ref{fig1},  the microstates with energy lower than $E$ are located inside a rectangle with sides $2\sqrt{2mE}$ and $L$. Since each microstates corresponds to  an area $h$, the Planck constant, we obtain
\begin{equation}
N_1^{(1)}(E)=\frac{2L}{h}\sqrt{2mE}\,,\qquad n_1^{(1)}(E)=\frac{L}{h}\sqrt{\frac{2m}{E}}\,,
\label{dens-1-1-ig}
\end{equation}
for the one-dimensional single-particle densities of states. 
Thus $C_1^{(1)}=\sqrt{2m}L/h$, $\kappa_1^{(1)}=1/2$ and  Eq.~(\ref{dens-1-D}) immediately gives the single-particle density of states in $D$ dimensions:
\begin{equation}
n_1^{(D)}(E)=V_D\left(\frac{2\pi m}{h^2}\right)^{D/2}\frac{E^{D/2-1}}{\Gamma(D/2)}\,.
\label{dens-1-D-ig}
\end{equation}
The microcanonical density of states of the  ideal gas with $N$ particles of mass $m$ in a volume $V_D$ follows from Eq.~(\ref{dens-N}) with $\kappa_1=D/2$, $C_1\Gamma(\kappa_1)=V_D(2\pi m/h^2)^{D/2}$, according to Eq.~(\ref{dens-1-D-ig}), so that
\begin{equation}
n_N^{(D)}(E)=V_D^N\left(\frac{2\pi m}{h^2}\right)^{ND/2}
\frac{E^{ND/2-1}}{\Gamma(ND/2)}
\label{dens-N-D-ig}
\end{equation}
and the integrated density of states is given by
\begin{equation}
N_N^{(D)}(E)=V_D^N\left(\frac{2\pi m}{h^2}\right)^{ND/2}
\frac{E^{ND/2}}{\Gamma(ND/2+1)}\,,
\label{int-dens-N-D-ig}
\end{equation}
in agreement with the result obtained by standard methods.\cite{kubo67a,huang87,mazenko00}
For undistinguishable particles the densities in Eqs~(\ref{dens-N-D-ig}) and~(\ref{int-dens-N-D-ig}) must be divided by $N!$, the number of permutations between the $N$ particles.

With an ideal gas  consisting of $N_a$ particles of mass $m_a$ and $N_b$ particles of mass $m_b$, such that $N_a+N_b=N$, Eq.~(\ref{dens-N-ab}) leads to
\begin{equation}
n^{(D)}_{N_a,N_b}(E)=V_D^N
\left(\frac{2\pi}{h^2}\right)^{ND/2}\left(m_a^{N_a} m_b^{N_b}\right)^{D/2}
\frac{E^{ND/2-1}}{\Gamma(ND/2)}
\label{dens-N-D-ig-ab}
\end{equation}
for the microcanonical density of states of the ideal gas mixture in $D$ dimensions. The integrated density of states is given by
\begin{equation}
N^{(D)}_{N_a,N_b}(E)\!=\!V_D^N
\left(\frac{2\pi}{h^2}\!\right)^{ND/2}\!\!\!\left(m_a^{N_a} m_b^{N_b}\!\right)^{D/2}\!\!\!
\frac{E^{ND/2}}{\Gamma(ND/2\!+\!1)},
\label{int-dens-N-D-ig-ab}
\end{equation}
in agreement with Ref.~\onlinecite{fernandez-pineda79}. A  division of the densities by $N_a!N_b!$ is needed when the two species consist of undistinguishable particles.

\subsection{Classical harmonic oscillators}

\begin{figure}[tbh]
\includegraphics[width=\columnwidth]{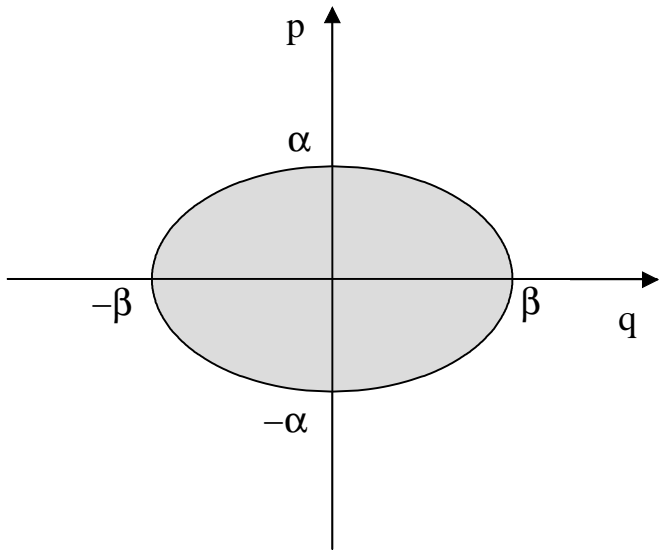}
\caption{Elliptic Domain of accessible microstates with energy lower than $E$ in the phase space of a one-dimensional harmonic oscillator with mass $m$ and angular frequency $\omega$. The semiaxes are $\alpha=\sqrt{2mE}$ and $\beta=\sqrt{2E/m}/\omega$. }
\label{fig2}
\end{figure}


We consider a collection of $N$ noninteracting classical harmonic oscillators in $D$ dimensions  which for simplicity are assumed to be isotropic.

For a single oscillator in one dimension the Hamiltonian reads
\begin{equation}
H(p,q)=\frac{p^2}{2m}+\frac{m\omega^2}{2}q^2\,.
\label{H-ho}
\end{equation}
The line of constant energy $E$ in the two-dimensional phase space is an ellipse with semiaxes $\alpha=\sqrt{2mE}$ and $\beta=\sqrt{2E/m}/\omega$ enclosing an area $\pi\alpha\beta=2\pi E/\omega$ shown in Fig.~\ref{fig2}. Each microstate corresponding to an area $h$, we obtain the integrated denstity of states as
\begin{equation}
N_1^{(1)}(E)=\frac{2\pi}{h\omega} E\,,\qquad n_1^{(1)}(E)=\frac{2\pi}{h\omega}\,.
\label{dens-1-1-ho}
\end{equation}
Thus $C_1^{(1)}=2\pi/h\omega$ and $\kappa_1^{(1)}=1$. According to  Eq.~(\ref{dens-1-D}) the density of states for  one oscillator in $D$ dimensions is
\begin{equation}
n_1^{(D)}(E)=\left(\frac{2\pi}{h\omega}\right)^D\frac{E^{D-1}}{(D-1)!}
\label{dens-1-D-ho}
\end{equation}
so that $\kappa_1=D$, $C_1\Gamma(\kappa_1)=(2\pi/h\omega)^D$, and, according to Eq.~(\ref{dens-N}), the microcanonical density of states for $N$ oscillators in $D$ dimensions reads
\begin{equation}
n_N^{(D)}(E)=\left(\frac{2\pi}{h\omega}\right)^{ND}\frac{E^{ND-1}}{(ND-1)!}
\label{dens-N-D-ho}
\end{equation}
and one obtains
\begin{equation}
N_N^{(D)}(E)=\left(\frac{2\pi}{h\omega}\right)^{ND}\frac{E^{ND}}{(ND)!}
\label{int-dens-N-D-ho}
\end{equation}
for the integrated density of states in agreement with the known result.\cite{beauregard65,kubo67b,fernandez-pineda79}
The generalization to a collection of anisotropic oscillators is immediate: It amounts to replace $\omega^D$ by $\prod_{i=1}^D\omega_i$.

\subsection{Ideal gas in a gravitational field}
We  consider a classical ideal gas with $N$ point particles of mass $m$ in $D$ dimensions. The gravitational field  with acceleration $g$ acts downwards along the vertical direction. The gas is confined inside a vessel with height $l$ in the vertical direction and  section $S_D=L^{D-1}$ in the transverse directions.  This is an example for which the single-particle density of states is {\it not } a simple power of $E$ so that we shall use the Laplace transform method.

\begin{figure}[tbh]
\includegraphics[width=\columnwidth]{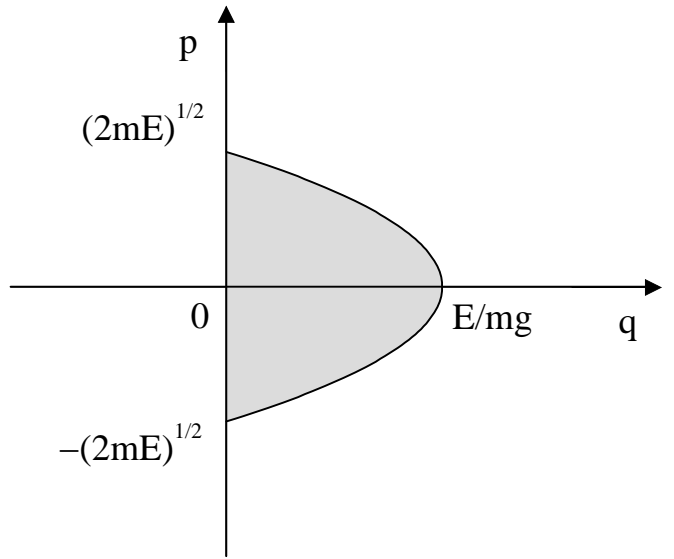}
\caption{Parabolic domain of accessible microstates with energy lower than $E$ in the phase space of a particle with mass $m$, moving in the one-dimensional half-space $q>0$, in a gravitational field with uniform acceleration $g$ directed downwards. When the motion is restricted to $0<q<l$, the domain of accessible microstates is restricted to the part of the shaded region with $q<l<E/mg$ when $E>mgl$.  }
\label{fig3}
\end{figure}


Let us calculate the single-particle density of states starting from Eq.~(\ref{conv-D-ab}) since the system is anistropic.

For each of the $D_a=D-1$ transverse dimensions, the single-particle density of states is  given by Eq.~(\ref{dens-1-1-ig}) so that $C_1^{(a)}=\sqrt{2m}L/h$, $\kappa_1^{(a)}=1/2$ and according to Eq.~(\ref{z1s})
\begin{equation}
z_1^{(a)}(s)=\frac{L\sqrt{2m}}{h}\frac{\Gamma(1/2)}{s^{1/2}}=\frac{L\sqrt{2\pi m}}{h\,s^{1/2}}\,.
\label{zas}
\end{equation}

In the remaining dimension ($D_b=1$) we have to calculate the one-dimensional  density of states for a particle in a gravitational field. The Hamiltonian reads
\begin{equation}
H(p,q)=\frac{p^2}{2m}+mgq\,,\qquad 0<q<l\,.
\label{H-gf}
\end{equation}
The line of constant energy in phase space is the segment of  parabola with equation $p=\pm\sqrt{2m}\sqrt{E-mgq}$ in the region $0<q<l$  (see Fig.~\ref{fig3}). Thus the number of microstates with energy lower than $E$ is obtained by dividing by $h$ the surface between the parabola and the $p$ axis (restricted to the region where $q<l$  when $E>mgl$)  so that
\begin{eqnarray}
\!\!\!\!\!\!\!\!\!\!N_1^{(b)}(E)=&&\!\!\!\!\!\!\frac{2\sqrt{2m}}{h}\!\!\int_0^{\min(\frac{E}{mg},l)}\sqrt{E-mgq}\,dq\nonumber\\
=&&\!\!\!\!\!\!\frac{2\sqrt{2m}}{hmg}\!\!\int_{\max(E-mgl,0)}^E\sqrt{t}\,dt\nonumber\\
=&&\!\!\!\!\!\!\frac{4\sqrt{2m}}{3hmg}\!\!\left[E^{3/2}\!\!-\!(\!E-mgl)^{3/2}\Theta(\!E\!-\!mgl)\right]\!,
\label{dens-int-1-gf}
\end{eqnarray}
where $\Theta(x)$ is the Heaviside step function such that  $\Theta(x)=0$ when $x<0$ and $\Theta(x)=1$ when $x\geq 0$.
The derivative with respect to $E$ leads to
\begin{equation}
n_1^{(b)}(E)=\frac{2\sqrt{2m}}{hmg}\!\left[E^{1/2}\!\!-\!(\!E\!-\!mgl)^{1/2}\Theta(\!E\!-\!mgl)\right].
\label{dens-1-gf}
\end{equation}
\begin{widetext}
Taking the Laplace transform of this density of states, one obtains
\begin{eqnarray}
z_1^{(b)}(s)&=&\frac{2\sqrt{2m}}{hmg}\left[\int_0^\infty E^{1/2}e^{-sE}\,dE
-e^{-smgl}\int_{mgl}^\infty( E-mgl)^{1/2}e^{-s(E-mgl)}dE\right]\nonumber\\
&=&\frac{2\sqrt{2m}}{hmg}\frac{\Gamma(3/2)}{s^{3/2}}\left(1-e^{-smgl}\right)=\frac{\sqrt{2\pi m}}{hmg}
\frac{\left(1-e^{-smgl}\right)}{s^{3/2}}\,.
\label{zbs}
\end{eqnarray}
The Laplace transform of the $N$-particle density of states in $D$ dimensions then follows from Eqs~(\ref{conv-N}) and~(\ref{conv-D-ab}) together with  Eqs~(\ref{zas}) and~(\ref{zbs})
\begin{eqnarray}
z_N^{(D-1,1)}(s)&=&[z_1^{(a)}(s)]^{N(D-1)}[z_1^{(b)}(s)]^{N}=\left(\frac{S_D}{mg}\right)^N\left(\frac{2\pi m}{h^2}\right)^{ND/2}\frac{\left(1-e^{-smgl}\right)^N}{s^{N(D+2)/2}}\nonumber\\
&=&\left(\frac{S_D}{mg}\right)^N\left(\frac{2\pi m}{h^2}\right)^{ND/2}\sum_{k=0}^N(-1)^k{N\choose k}\frac{e^{-skmgl}}{s^{N(D+2)/2}}\,.
\label{zDNs}
\end{eqnarray}
The inverse Laplace transform is obtained using the transformation formula
\begin{equation}
\frac{e^{-sa}}{s^\mu}\longleftrightarrow\frac{(E-a)^{\mu-1}}{\Gamma(\mu)}\,\Theta(E-a)\,,\qquad (\mu>0)\,.
\label{Lt}
\end{equation}
 for each term in the sum. It leads to the following expression for the density of states
\begin{equation}
n_N^{(D-1,1)}(E)\!=\!\left(\frac{S_D}{mg}\right)^N\!\!
\left(\frac{2\pi m}{h^2}\right)^{ND/2}\sum_{k=0}^N(-1)^k{N\choose k}
\frac{(E-kmgl)^{N(D+2)/2-1}}{\Gamma[N(D+2)/2]}\,\Theta(E-kmgl)\,,
\label{dens-ND-gf}
\end{equation}
and the integrated density of states is given by
\begin{equation}
N_N^{(D-1,1)}(E)\!=\!\left(\frac{S_D}{mg}\right)^N\!\!
\left(\frac{2\pi m}{h^2}\right)^{ND/2}\sum_{k=0}^N(-1)^k{N\choose k}
\frac{(E-kmgl)^{N(D+2)/2}}{\Gamma[N(D+2)/2+1]}\,\Theta(E-kmgl)\,,
\label{dens-int-ND-gf}
\end{equation}
in agreement with the result of Ref.~\onlinecite{roman97}. 
\end{widetext}

When $l\to\infty$, only the first term  contributes to the sum and one recovers the result of Ref.~\onlinecite{roman95}. This result  can be obtained more directly since the deviation of the single-particle density of states from a simple power law coming from the last term in Eq.~(\ref{dens-1-gf}) disappears  in this limit. 

Here too, one has to divide the densities by $N!$ when undistinguishable particles are considered.

\section{Summary and conclusion}
We have shown how, using the integral equation~(\ref{rec-dens}),  one can construct the $N$-particle density of states  iteratively,  starting from the single-particle density of states and, in the same way, using the integral equation~(\ref{rec-dens-1}), how to deduce  the single-particle density of states in $D$ dimensions from the single-particle density of states in one dimension  for a microcanonical system of noninteracting particles with a separable Hamiltonian. 

The solutions of the integral equations are easily obtained when the single-particle density of states in one dimension follows a power law. This has been illustrated on two classical examples: The Boltzmann gas and a collection of  classical harmonic oscillators. In other cases, the  density of states may be obtained using a Laplace transform method. Since the Laplace transform of the microcanonical density of states is given by the canonical partition function of the system, it factorizes for noninteracting particles. As a consequence it is easily deduced from the form of the single-particle densities of states for each dimension as shown for an ideal gas in a vessel of finite height under the influence of the gravitational field. 

\begin{acknowledgments}
It is a pleasure to thank Dragi Karevski for helpful discussions. 
\end{acknowledgments}

\end{document}